\documentclass[mathleft
]{an}
\usepackage{graphicx}
\usepackage{times}
\begin{document}

\Pagespan{1}{}
\Yearpublication{2009}%
\Yearsubmission{2009}%
\Month{xx}%
\Volume{xxx}%
\Issue{xx}%

\arraycolsep0.35mm                      
\catcode`\@=11
\def\gsim{\ifmmode{\,\mathrel{\mathpalette\@versim>\,}}
    \else{$\,\mathrel{\mathpalette\@versim>}\,$}\fi}
\def\lsim{\ifmmode{\,\mathrel{\mathpalette\@versim<\,}}
    \else{$\,\mathrel{\mathpalette\@versim<}\,$}\fi}
\def\@versim#1#2{\lower 2.9truept \vbox{\baselineskip 0pt \lineskip
    0.5truept \ialign{$\m@th#1\hfil##\hfil$\crcr#2\crcr\sim\crcr}}}
\catcode`\@=12  
\def\GHz{{\rm GHz}}
\def\Msol{M_{\odot}}
\def\Mstar{{M}_{*}}
\def\Lstar{{L}_{*}}
\def\pn{\par\noindent}
\def\bs{\bigskip}
\def\ms{\medskip}
\def\ss{\smallskip}
\def\M{{M}}
\def\Msun{{M}_{\odot}}
\let\msun=\Msun
\def\Tvir{T_{\rm vir}}
\def\Mcrit{{M}_{\rm crit}}
\def\Mmin{{M}_{\rm min}}
\newcommand{\ls}{{_<\atop^{\sim}}}
\newcommand{\lax}{{_<\atop^{\sim}}}
\newcommand{\gs}{{_>\atop^{\sim}}}
\newcommand{\gax}{{_>\atop^{\sim}}}

\title{Cool gas accretion, thermal evaporation and quenching of star formation in elliptical galaxies}

\author{C. Nipoti\thanks{Corresponding author:
  \email{carlo.nipoti@unibo.it}\newline}
}
\titlerunning{Thermal evaporation in elliptical galaxies}
\authorrunning{C. Nipoti}
\institute{
Dipartimento di Astronomia, Universit\`a di Bologna, via Ranzani 1, I-40127 Bologna, Italy}

\date{13 August 2009}


\abstract{The most evident features of colour-magnitude diagrams of
  galaxies are the red sequence of quiescent galaxies, extending up to
  the brightest elliptical galaxies, and the blue cloud of
  star-forming galaxies, which is truncated at a luminosity $L\sim
  \Lstar$. The truncation of the blue cloud indicates that in the most
  massive systems star formation must be quenched.  For this to happen
  the virial-temperature galactic gas must be kept hot and any
  accreted cold gas must be heated. The elimination of accreted cold
  gas can be due to thermal evaporation by the hot interstellar
  medium, which in turn is prevented from cooling by feedback from
  active galactic nuclei.}
\maketitle

\section{Need for quenching of star formation in galaxies}

Colour-magnitude diagrams of galaxies are dominated by the red
sequence of quiescent galaxies and the blue cloud of star-forming
galaxies (Blanton et al. 2003, Baldry et al. 2004), with a green
valley in between, in which a minority of galaxies lie (Driver et
al. 2006). While the red sequence extends up to the bright end of the
galaxy distribution, the blue cloud is truncated at a luminosity
$L\sim \Lstar$, so there are not very massive blue, star-forming
galaxies.  Such a feature can be reproduced in galaxy formation models
only assuming that star formation is effectively quenched in the
galaxies with the deepest potential wells (e.g.  Bower et al. 2006;
Croton et al.  2006; Cattaneo et al.~2008).  What are the processes
responsible for this quenching is still matter of debate.

For star formation to cease in a galactic system, it is necessary that
(1) hot (virial-temperature) gas does not cool efficiently and (2)
accreted cool gas is heated before it can form stars (Binney 2004).
Task (1) can be accomplished by feedback from Active Galactic Nuclei
(AGN). Though the details of how AGN feedback works are controversial,
empirical evidence that this mechanism is effective comes from studies
of cool cores in galaxy clusters (Birzan et al. 2004; McNamara \&
Nulsen 2007), suggesting that also in massive galaxies
virial-temperature gas is kept hot by the intermittent jets of the
central radio source (Binney 2004; Nipoti \& Binney 2005).  Other
mechanisms, such as gravitational heating by clumpy accretion (Dekel
\& Birnboim 2008, Khochfar \& Ostriker 2008), can contribute to
prevent virial-temperature gas from cooling, but non-gravitational
heating is necessary to drive gas out of the galaxy potential well and
thus explain, for instance, why the intracluster medium is
metal-enriched (Renzini~1997). Task (2) is not less important than
task (1), because there is growing evidence that cold gas accretion is
a common phenomenon during the galaxy lifetimes: the strongest
evidence for accretion comes from observation of late-type galaxies
(Sancisi et al. 2008), but cold gas structures are observed also
around early-type galaxies (Morganti et al. 2006).  Nevertheless, not
so much attention has been devoted to finding the mechanism
responsible for eliminating recently accreted cool gas. Here we
briefly review the proposal that this task can be accomplished by
ablation and thermal evaporation of accreted cold gas by
virial-temperature gas (Binney 2004; Nipoti \& Binney 2004, 2007).
 
\section{Quenching by thermal evaporation in $L>\Lstar$
  elliptical galaxies}

An important difference between very massive elliptical galaxies and
less massive galaxies is that the former are hot-gas rich, in the
sense that they contain large amounts of relatively dense, X-ray
emitting, virial-temperature gas, while the latter are hot-gas poor,
in the sense that their diffuse X-ray emission is hardly detectable
(lower mass elliptical galaxies; e.g. David et al. 2006) or not
detected at all (disc galaxies; e.g. Rasmussen et al. 2009). This
means that in lower-mass galaxies the virial-temperature gas halos,
which are expected to exist on theoretical grounds, are very rarefied.

These observational findings are reasonably well understood
theoretically, in terms of the existence of a critical dark-matter
halo mass $\Mcrit\sim 10^{12}\Msun$.  Only in halos with mass above
$\Mcrit$ a significant fraction of the primordial infalling gas is
shock heated to the virial temperature (Binney 1977; Birnboim \& Dekel
2003; Kere{\v s} et al. 2005; Dekel \& Birnboim 2006) and the gravitational
potential wells are deep enough to retain gas heated by supernova
feedback (Dekel \& Silk~1986). As a consequence, denser and denser
hot-gas atmospheres build up only in systems with mass $M \gsim
\Mcrit$, while $M\lsim \Mcrit$ systems are expected to have very
rarefied coronae of virial-temperature gas.

Let us consider the process of cool gas accretion onto galaxies as a
function of the galaxy total mass $M$ or virial temperature $\Tvir$.
Infalling cool ($T\ll \Tvir$) clouds find very different physical
conditions, depending on whether the accreting galaxy has total mass
higher or lower than $\Mcrit$.  If $M \gsim \Mcrit$ the cool clouds
are likely to be eliminated by ablation and thermal evaporation
because of the high temperature and relatively high density of the hot
interstellar medium (Binney 2004; Nipoti \& Binney 2004), while cool
clouds are likely to survive if $M \lsim \Mcrit$.  Nipoti \& Binney
(2007) explored this problem quantitatively, by calculating the
minimum rate of ablation with a simple model based on analytic
estimates of the evaporation rate (Cowie \& McKee 1977; Cowie \&
Songaila 1977; McKee \& Cowie 1977). Cool gas clouds less massive than
a minimum mass $\Mmin$ are evaporated by thermal conduction before
they can form stars.  Though the estimate of $\Mmin$ is affected by
the uncertainties on the suppression of thermal conduction by tangled
magnetic fields, the ratio of $\Mmin$ in different systems is a robust
quantity. The minimum mass of clouds that can survive evaporation in a
representative hot-gas rich, very massive ($M> \Mcrit$) galaxy is a
factor of $\sim 1000$ larger than in a representative hot-gas poor,
less massive ($M< \Mcrit$) galaxy, even though the mass ratio between
the galaxies is just a factor of $10$.  As a consequence, the
aggregate mass of gas available for star formation, \emph{per unit
  galaxy mass}, is a factor of $\sim 10$ larger in the low-mass system
than in the high-mass system (see Nipoti \& Binney 2007 for
details). The bottom line is that thermal evaporation of cool clouds
by the hot interstellar medium can give an important contribution to
quench star formation in the most massive elliptical galaxies
($L>\Lstar$).

\section{What about $L< \Lstar$ galaxies?}

We have seen that thermal evaporation can explain the truncation of
the blue cloud at $L\gsim \Lstar$.  At luminosities $L\lsim \Lstar$ we
find galaxies in both the red sequence and the blue cloud.  All these
systems have galactic dark-matter halos with masses $\lsim \Mcrit$,
relatively low virial temperature and quite rarefied hot-gas
atmospheres. Therefore, according to the results of Nipoti \& Binney
(2007), thermal evaporation is not efficient\footnote{This does not
  imply that thermal conduction is unimportant in $M\lsim \Mcrit$
  systems: combined with buoyancy, even relatively strongly suppressed
  thermal conduction can stabilize the galactic coronae against
  thermal instability (Binney, Nipoti \& Fraternali 2009).} and star
formation can proceed as long as cool gas is accreted.

So what determines whether a $L< \Lstar$ galaxy lies in the blue cloud
or in the red sequence?  It seems likely that the key factor is
environment: the fraction of galaxies in the red sequence is observed
to increase for increasing density of the environment (from voids to
clusters), and the effect is stronger for lower-mass galaxies (Baldry
et al. 2006).  This trend can be explained by considering that a
$L<\Lstar$ galaxy belonging to a big group or a cluster has little
chance of accreting cool gas, which can be easily eliminated via a
combination of ram-pressure stripping and thermal evaporation by the
hot intragroup or intracluster medium. However, in contrast with the
brightest elliptical galaxies, it is not excluded that $L<\Lstar$
elliptical galaxies have experienced small recent episodes of star
formation, if they happened to accrete some cool gas.  This accounts
for the fact that lower-luminosity ellipticals have cuspier central
luminosity profiles and younger central stellar populations than the
most luminous ellipticals (Nipoti \& Binney 2007; Nipoti 2009).

\end{document}